 \documentstyle [12pt]{article}
  \newcommand{\be}{\begin{equation}}
  \newcommand{\ee}{\end{equation}}
  \newcommand{\bdis}{\begin{displaymath}}
  \newcommand{\edis}{\end{displaymath}}

  \title{Helical shell models for three dimensional turbulence.}
  \author{ R. Benzi$^{1}$, L. Biferale$^{1}$ and E.Trovatore$^{2}$ }
  
\begin{document}
  \maketitle
  \centerline{$^{1}$  Dipartimento di Fisica, Universit\`{a}
di Tor Vergata}
  \centerline{Via della Ricerca Scientifica 1, I-00133 Roma, Italy.}
  \centerline{$^{2}$  INFM-Dipartimento di Fisica, Universit\`{a}
di Cagliari}
  \centerline{Via Ospedale 72, I-09124, Cagliari, Italy.}
  \date{ }
  \medskip

  \begin{abstract}
  In this paper we study a new class of shell models, defined
in terms of two complex
  dynamical
  variables per shell,  transporting positive and negative helicity
  respectively.\\
  The dynamical equations are
  derived from   a
  decomposition into
  helical modes of the velocity Fourier components of Navier-Stokes
  equations (F. Waleffe, Phys. Fluids A {\bf 4}, 350 (1992)).

  This decomposition  leads to four different types
  of shell models, according to the possible non-equivalent
  combinations of  helicities
  of the three interacting modes in each triad.\\
  Free parameters are fixed by
  imposing the conservation of  energy  and of a ``generalized helicity''
  $H_\alpha$ in the inviscid and unforced limit.

  For $\alpha=1$ this non-positive invariant looks exactly like helicity
  in the Fourier-helical decomposition of the Navier-Stokes equations.\\
   Long numerical integrations are performed, allowing the
computation of the
  scaling exponents  of the velocity increments and energy flux moments.\\
  The dependence of the models on the generalized helicity
parameter $\alpha$
  and on the scale parameter $\lambda$ is also studied.\\
PDEs are finally derived in the limit when
the ratio between shells goes to one.

   \end{abstract}

  \newpage

  \section{Introduction.}

The most intriguing problem in 3 dimensional turbulence
 is related to the understanding of the dynamical mechanism triggering
 and supporting the energy
 cascade from large to small scales. According to the celebrated
Kolmogorov
 theory \cite{K} (K41), energy should be transferred downwards in scales
 following a  self-similar and  homogeneous process. This idea,
 plus the assumption of local isotropy and small scales universality,
 led Kolmogorov to formulate a precise prediction on the scaling
 properties of the increments of turbulent velocity fields,
$\delta v (l) \sim
 |v(x+l)-v(x)|$, at
 distances $l$.
    Namely, following Kolmogorov and denoting
 by $\epsilon(l)$ the energy transfer through scale $l$, we have:

\be
S_p(l) \equiv \langle(\delta v(l))^p\rangle=C_p
\langle (\epsilon(l))^{p/3} \rangle l^{p/3}
\sim l^{\zeta(p)}
\label{eq.1}
\ee
where $C_p$ are constants and the scale $l$ is supposed to be in
the inertial range,
i.e. much smaller than the integral scale
  and much larger than the viscous
 dissipation cutoff.
{}From (\ref{eq.1}) we deduce that depending on the scaling properties of
the energy transfer, we can observe different power law behaviors
for the structure
functions $S_p(l)$. Kolmogorov made the simplest assumption by
taking $\epsilon(l)$
to be constant and  deducing that:
\be
\zeta(p) = p/3, \forall p
\ee
While from a qualitative point of view the Kolmogorov intuition was a true
breakthrough in the understanding of turbulence, his theory
lacks quantitative
agreement with experiments. In particular, there are many
experimental and numerical \cite{MS,BCTBS,BCBC}
results telling  us that energy
is transferred intermittently, leading to non-trivial scaling corrections
to the ``p over 3'' Kolmogorov prediction for the $\zeta(p)$ exponents.

More and more efforts have been devoted  in trying to
 understand and
 to model with
good accuracy the energy transfer dynamics. \\
Beside analytical and direct numerical approach, there are two other
 possible choices:
either building simple random processes for
the chaotic energy transfer among different scales \cite{BPPV,MSbis,SL,D}
 or studying a
dynamical deterministic model. \\
 In the last $20$ years, among the possible
dynamical modelizations of fully developed energy transfer, those obtained
by using {\it shell models} have been particularly successfully.
These models
concentrate only on the dynamical interactions among degrees of
freedom at different scales, neglecting completely
their spatial location. In this way a one-dimensional chain of
interacting Fourier modes is constructed.
The same
non-linear structure of Navier-Stokes eqs. is retained, but their
three-dimensional vectorial nature is completely lost.
The simplifications are so strong that only
{\it a posteriori} one can say whether the model is interesting
and reliable or not.

 For the most famous among all shell models, the Gledzer-Ohkitani-Yamada
   (GOY) model, are now available a wide number of works and contributes
    (\cite{G}-\cite{BK}). The intermittent and chaotic dynamics of the
GOY model is remarkably close (for a suitable  choice of the free
parameters)  to what is found  for the Navier-Stokes
eqs.

The most striking property of the GOY model
    is that, for some particular choices of the free parameters, it has
    an intermittent and chaotic dynamics quantitatively close to what
    is found for the Navier-Stokes equations. In particular,
    the set of scaling exponents $\zeta(p)$ coincides (for a suitable
choice
    of parameters) with that measured in true turbulent
    flows.

   Recently, it has been pointed out that the
   GOY  capability of having a chaotic and intermittent behavior with
   scaling exponents very close to the real ones, is strictly related to
    the conservation in the inviscid, unforced limit
     of two quadratic quantities. The first quantity is
    the {\it{energy}}, while the second is, roughly speaking,
    the equivalent of {\it{helicity}} in 3D turbulence
   \cite{BKLMW}.

   Both the {\it{GOY-helicity}} \cite{BK} and the
    helicity in the N-S equations \cite{AL,PS}
     have been suggested to play
     an important role in
    triggering the intermittent nature of the energy cascade.

   This new emphasis on the role played by the helicity
    motivated  us to study  a new class
    of shell models. These models can be seen as a generalization
    of the GOY model such that their helical structure is
    closer to the corresponding N-S case (indeed, one of them
    is nothing but the old GOY case).

    We study shell models with two complex variables per shell,
    derived on the basis of an
    helical formulation of the N-S equations in the Fourier space.

    In this way, it is possible to obtain
    a second non-positive defined invariant
     closer to the definition of
    helicity in the N-S equations. Our aim consists
in trying to understand
    the  importance of both (competing?) transfer of energy and helicity
    in N-S eqs. by examining the non trivial dynamics shown by
this new class
    of helical-shell models.

    The outline of the paper is the following:
    in section 2 we review the GOY model;
    in section 3 we introduce the new class of
    helical-shell models; in section 4 the basic  triad interactions
     within
    three contiguous shells are studied. Section 5 contains  the results
     of  our numerical
    simulations; in section 6  PDEs for the continuum limit (ratio
    between shells that goes to one)  is derived;
     conclusions follow  in section 7.

\section{The GOY model.}

 The GOY model can be seen as a severe truncation of the
Navier-Stokes equations:
  it retains only one complex mode $u_n$ as a representative
of all Fourier modes in
 the shell of wave numbers $k$ between $k_n=k_0\lambda^n$
and $k_{n+1}$, $\lambda$
  being an arbitrary scale parameter ($\lambda>1$), usually
taken equal to $2$.

The  dynamics is governed by the following set of complex coupled ODEs
where only couplings with the nearest and next nearest shells are kept:

  \be
  \frac{d}{dt} u_n =i\, k_n \left(a u^{*}_{n+1}u^{*}_{n+2} +
  b u^{*}_{n+1}u^{*}_{n-1}
  +c u^{*}_{n-1}u^{*}_{n-2} \right)
   -\nu k_n^2 u_n +\delta_{n,n_0}f,
  \ee

where $\nu$ is the viscosity, $f$ the external forcing acting
on a large scale $n_0$ and $a$,$b$,$c$ are three free parameters.
By adjusting the time scale we can always fix $a=1$; the possible
choices for $b$,$c$
are restricted by imposing
 the conservation of two quadratic quantities in the inviscid
and unforced limit:
  \be
  W_{1,2} = \sum_n
   z_{1,2}^n |u_n|^2,
  \ee

where $z_{1,2}$ are the solutions of the quadratic equation:
\be
c \lambda^2 z^2+b \lambda z+a=0.
\ee

In order to stay as close as possible to the N-S equations,
we require that
one of the two conserved quantities is the energy, i.e. $z_1=1$:
  \be
  W_1=E = \sum_n |u_n|^2.
  \ee

If we rewrite:

\be
\begin{array}{ll}
\lambda b = -\epsilon,\\
\lambda^2 c = \epsilon-1,\\
\end{array}
\label{eq:epsilon}
\ee

we are left with only two ($\lambda$ and $\epsilon$) of the original four
free parameters.
 The second quadratic invariant is:
  \be
  W_2=H = \sum_n
   (\epsilon-1)^{-n} |u_n|^2.
  \label{eq:H}
  \ee

  The characteristics of this second invariant change by
changing $\epsilon$:
  when $\epsilon<1$ it is not positive-defined (as helicity in 3D),
 while if $\epsilon>1$
  it is positive defined (as enstrophy in 2D).
  Expression (\ref{eq:H}) can be rewritten as:

  \be
  H_\alpha = \sum_n
   \chi(\epsilon)^{n} k^{\alpha(\epsilon,\lambda)}_n |u_n|^2,
  \label{eq:Halpha}
  \ee

where $\chi(\epsilon)=sign(\epsilon-1)$ and the $\alpha$
parameter is related
to $\epsilon$ and $\lambda$ by:
\be
\vert \epsilon-1 \vert = \lambda^{-\alpha}.
\label{eq:alpha}
\ee

Our interest here is to consider how well the dynamics of a 3D turbulent
 flow is reproduced by the model: only $\epsilon$ in the range
  $0 < \epsilon < 1$ will be taken, in order to have a non-positive
   defined second invariant.
Indeed, for $\alpha=1$ our ``generalized helicity" $H_\alpha = \sum_n
   (-1)^{n} k^\alpha_n |u_n|^2$ has physical dimensions coinciding
with the 3D
Navier-Stokes helicity.
The two free parameters of the model can be taken to be $\lambda$
(the ratio between adjacent shells) and $\alpha$.
The two coefficients $b$ and $c$ can be rewritten as:
\be
\begin{array}{ll}
b = \lambda^{-\alpha-1}-\lambda^{-1},\\
c = -\lambda^{-\alpha-2}.
\end{array}
\ee

Such a class of models has a highly non trivial dynamical
behavior. Intermittency of the energy transfer and multifractal
nature of energy dissipation
have been studied in \cite{G,YO,JPV,BLLP,PBCFV}.

It turns out that the values of the $\zeta(p)$s are not
universal, depending
on the choice of $\epsilon$ and $\lambda$ \cite{BLLP,BKLMW}.
Nevertheless, Kadanoff {\it{et al.}} \cite{BKLMW} verified that the
 scaling exponents
are invariant along the curve in the $(\epsilon,\lambda)$ plane where
both energy and helicity are conserved, i.e. the curve at
$\alpha(\epsilon,\lambda)=1$.

This suggested that the second invariant plays a crucial role in the
 model dynamics.

More recently Biferale and Kerr \cite{BK} have attributed to the helicity
 the role
 of triggering the intermittent cascade of the energy from large
to small scales.

 These considerations together with the observation that
this ``GOY-helicity''
 is only partially consistent
 with the N-S helicity (i.e. it presents an asymmetry between odd and
 even shells that does not have any counterpart in physical flows)
  persuaded us to study a modified shell model \cite{BK}, with two complex
  variables in each shell, carrying helicity of opposite sign,
in order to obtain
   a second invariant closer
   to the N-S helicity.
In the following section we introduce this new class of shell models,
whose nonlinear
 interactions are constructed on the basis of a helical
decomposition of the
 N-S equations in the Fourier space.

\section{The helical-shell models.}

In order to introduce two helical variables per  shell
 we refer to the velocity field in N-S equations, expanded in a Fourier
  series \cite{W}.
The velocity vector can be represented in terms of its projection on an
orthogonal basis formed by $\bf{k}$, $\bf{h}_+$ and $\bf{h}_-$.
The two basis vectors $\bf{h}_+$ and $\bf{h}_-$ can be
chosen to be the eigenmodes of the curl
 operator:
\be
i {\bf{k}} \times {\bf{h_s}} = s k {\bf{h_s}},
\ee

where $s=\pm 1$.

This corresponds to an expansion of the velocity vector into helical modes:

\be
{\bf{u}}({\bf{x}})= \sum_{{\bf{k}}} {\bf{u}}({\bf{k}})
\exp({i {\bf{k \cdot x}}})=
\sum_{{\bf{k}}} [u^+({\bf{k}}) {\bf{h}}_+ + u^-({\bf{k}}) {\bf{h}}_-]
\exp({i {\bf{k \cdot x}}}).
\label{eq:Fourier}
\ee

The real flow velocity corresponding to the plus (minus) mode rotates
clockwise (counterclockwise) as one moves in the direction of $\bf{k}$,
 thereby forming a left-handed (right-handed) helix; the vorticity vector
 of such a flow is parallel (antiparallel) to the velocity and the helicity
  is maximum (minimum).

The kinetic energy and helicity are given by:

\be
\begin{array}{ll}
E= \sum_{{\bf{k}}} E({\bf{k}})=
 \sum_{{\bf{k}}} \frac{1}{2}
{\bf{u}}({\bf{k}}) \cdot {\bf{u}}^*({\bf{k}}) =
  \sum_{{\bf{k}}} (\vert {\bf{u}}^+({\bf{k}})
 \vert ^2 + \vert {\bf{u}}^-({\bf{k}}) \vert ^2), \\
H= \sum_{{\bf{k}}} H({\bf{k}})=
 \sum_{{\bf{k}}} \frac{1}{2} {\bf{u}}({\bf{k}})
 \cdot {\bf{\omega}}^*({\bf{k}}) =
  \sum_{{\bf{k}}} k (\vert {\bf{u}}^+({\bf{k}})
\vert^2 - \vert {\bf{u}}^-({\bf{k}}) \vert^2 ).
\end{array}
\ee

Plugging eq.(\ref{eq:Fourier}) into the N-S equations yields to
the dynamical evolution for the
complex amplitudes $u^{s_k}({\bf{k}},t)$ $(s_k=\pm 1)$ \cite{W}:
\be
\frac{d}{dt} u^{s_k}({\bf{k}}) + \nu k^2 u^{s_k}({\bf{k}})
 = \sum_{{\bf{k+p+q}}=0} \sum_{s_p,s_q} g_{{\bf{k,p,q}}}
(s_p p - s_q q) (u^{s_p}({\bf{p}}) u^{s_q}({\bf{q}}))^*.
\ee

The geometric factor
$g_{{\bf{k,p,q}}}=- \frac{1}{4} ( {\bf{h_{s_k}
\wedge h_{s_p} \cdot h_{s_q}}})^*$
 can be developed and factorized:
\be
g=r \frac{s_k k +s_p p+ s_q q}{p},
\ee

where $r$ is a function of the triad shape only \cite{W}.

Eight different types of interaction between three modes
$u^{s_k}({\bf{k}})$,
 $u^{s_p}({\bf{p}})$, $u^{s_q}({\bf{q}})$
 with $|\bf{k}| < |\bf{p}| < |\bf{q}|$ are
  allowed according to the value of the triplet
$(s_k, s_p, s_q)$ =
  $(\pm 1,\pm 1,\pm 1)$:
among them, only four are independent, the coefficients of the
  interaction with reversed helicities $(-s_k, -s_p, -s_q)$
  being identical to those with
  $(s_k, s_p, s_q)$ \cite{W}:

  \begin{enumerate}
  \item  $(s_k, s_p, s_q)= (+,-,+)$ or $(-,+,-)$,
  \item  $(s_k, s_p, s_q)= (+,-,-)$ or $(-,+,+)$,
  \item  $(s_k, s_p, s_q)= (+,+,-)$ or $(-,-,+)$,
  \item  $(s_k, s_p, s_q)= (+,+,+)$ or $(-,-,-)$.
  \end{enumerate}

The equations corresponding to the single interaction $(s_k, s_p, s_q)$
have the form (omitting viscosity and forcing):
\be
\begin{array}{lll}
\dot{u}^{s_k}=r (s_p p-s_q q)
\frac{s_k k +s_p p+ s_q q}{p} (u^{s_p} u^{s_q})^*,\\
\dot{u}^{s_p}=r (s_q q-s_k k)
\frac{s_k k +s_p p+ s_q q}{p} (u^{s_q} u^{s_k})^*,\\
\dot{u}^{s_q}=r (s_k k-s_p p)
\frac{s_k k +s_p p+ s_q q}{p} (u^{s_k} u^{s_p})^*.
\end{array}
\label{eq:triad}
\ee

Each interaction independently conserves both energy and helicity on a
single triad.

The dynamical system (\ref{eq:triad}) composed by a single
triad  can be considered as the
basic brick of the semi-infinite chain leading to the
transfer of energy in turbulent flow.

By studying its stability properties it
is possible to understand how energy and helicity are transferred
among different wave-vectors belonging to the same triad. \\
Following \cite{W}, we distinguish two different kinds of dynamics:
for the cases corresponding to the choices 1 and 3 of the
triad helicities,
the unstable wave-vector is the smallest-one,
while for the cases 2 and 4 the  unstable wave-vector is
the medium-one. This
very simple analysis suggests that by linking together a series of
triads we  should have a forward energy transfer  for cases
1 and 3 and both forward and backward (competing)
energy transfers for cases 2 and 4.\\
 In a turbulent flow the direction
of energy transfer is dynamically controlled by the
triple correlation
$<u^{s_k}({\bf{k}})u^{s_p}({\bf{p}})u^{s_q}({\bf{q}})>$.
It is reasonable to argue that the statistical
properties of $<u^{s_k}({\bf{k}})u^{s_p}({\bf{p}})u^{s_q}({\bf{q}})>$ are
such that the overall direction in
energy transfer coincides with  the simplified behavior inferred from
the stability study of the single triad ({\it{instability assumption}} in
\cite{W}).
For instance, it is easy to estimate, by using the
{\it{instability assumption}},
what would be the net energy transfer in the above four cases if the
energy spectrum had the Kolmogorov scaling, $E(k) = k^{-5/3}$ \cite{W}:
\begin{itemize}
\item  1 and 3: direct energy cascade from large to small scales,
\item 4: reverse energy cascade from small to large scales,
\item 2: direct (reverse) energy cascade for local (nonlocal) triads.
\end{itemize}

The helical decomposition of the N-S eqs. suggested us the opportunity
of defining a different GOY-like shell model for each one of the above four
classes.  In each
shell we will have two complex dynamical variables $u^+_n$ and $u^-_n$,
transporting
  positive and negative helicity respectively:
\be
\begin{array}{ll}
\dot{u}^+_n=i k_n (a_j u^{s_1}_{n+2} u^{s_2}_{n+1}+b_j
u^{s_3}_{n+1} u^{s_4}_{n-1}
+c_j u^{s_5}_{n-1} u^{s_6}_{n-2})^*-\nu k^2_n u^+_n +\delta_{n,n_0} f^+,\\
\dot{u}^-_n=i k_n (a_j u^{-s_1}_{n+2}
u^{-s_2}_{n+1}+b_j u^{-s_3}_{n+1} u^{-s_4}_{n-1}
+c_j u^{-s_5}_{n-1} u^{-s_6}_{n-2})^*-\nu k^2_n u^-_n +\delta_{n,n_0} f^-,
\end{array}
\label{eq:shells}
\ee

where $j=1,...,4$ labels the
four different models and the helicity indeces in the
 non linear interactions are easily found
 for each of the four cases (see table(1)).

The coefficients $a_j,b_j,c_j$ are determined imposing, as
usually, the energy
 conservation:

\be
\frac{d}{dt} E=\frac{d}{dt}(\sum_n(\vert u^+_n \vert^2 +
\vert u^-_n \vert^2))=0,
\ee

that leads to the same relation for all models:

\be
a_j+b_j \lambda+c_j \lambda^2=0
\ee

By imposing also the conservation of the generalized helicity:

\be
\frac{d}{dt} H_\alpha=\frac{d}{dt}\sum_n k^\alpha_n
(\vert u^+_n \vert^2 - \vert u^-_n \vert^2)=0
\ee

we obtain different relations for the four models:

\begin{enumerate}
\item  $a_1-\lambda^{\alpha+1}b_1+\lambda^{2(\alpha+1)}c_1=0$,
\item  $a_2-\lambda^{\alpha+1}b_2-\lambda^{2(\alpha+1)}c_2=0$,
\item  $a_3+\lambda^{\alpha+1}b_3-\lambda^{2(\alpha+1)}c_3=0$,
\item  $a_4+\lambda^{\alpha+1}b_4+\lambda^{2(\alpha+1)}c_4=0$.
\end{enumerate}

Fixing $a_j=1$ one then finds the expressions for the
coefficients $b_j$ and
 $c_j$ in terms of the parameters
$\lambda$ and $\alpha$ (see table(2)).

Let us remark two important facts. First, model 1 is
nothing but two masked and
uncorrelated versions of the
 original GOY model, with dynamical variables
$(u^+_1,u^-_2,u^+_3,...)$ and
 $(u^-_1,u^+_2,u^-_3,...)$ respectively; rewriting the
coefficients $b_1$ and
  $c_1$ in terms of the usual parameters $\lambda$ and $\epsilon$ one
   can easily recover the standard GOY model coefficients.
 Second, also model 4 is formed by two independent
  sets of variables $(u^+_1,u^+_2,u^+_3,...)$ and
$(u^-_1,u^-_2,u^-_3,...)$,
  each of them conserving separately a positive-definite
quantity similar to
  enstrophy in 2D. Thus, model 4
is equivalent to   two
 uncorrelated GOY models for 2D turbulence \cite{FDB,AFS}.

The fact that the model 1 is formed by two uncorrelated GOY models is clearly
 due to our choice of taking only first
 and second-neighbor interactions.
Model 4, on the other hand, will always be the sum
 of two separated models for any choice
of the interacting modes composing the triads.

In the following, we
will refer to the  properties of model 1 intending
 corresponding properties of the GOY model.
 Model 4 will be studied only for
completeness.

\section{One-triad systems.}

Following the {\it{instability assumption}} \cite{W} that
connects the single triad
dynamics with the global statistical behavior of a
multi-triads flow, we repeat
the analog stability study for the three shells, single triad system.

By isolating three shells of waves numbers $k_{1}, k_2, k_{3}$, we
 can inspect their dynamical properties as determined by their mutual
  interactions.

For the positive-helicity modes we have:

\be
\begin{array}{lll}
\dot{u}^+_{1}=i k_{1} (u^{s_1}_{3} u^{s_2}_{2})^*,\\
\dot{u}^+_{2}=i k_{2} b_j (u^{s_3}_{3} u^{s_4}_{1})^*,\\
\dot{u}^+_{3}=i k_{3} c_j (u^{s_5}_{2} u^{s_6}_{1})^*,
\end{array}
\label{eq:triadbis}
\ee

where $j=1,...,4$ for the four models.
An analogous set of equations holds for the negative-helicity modes,
changing the sign of the helicity index of all variables.

This system conserves both energy and helicity.

The corresponding equations for the energies are the following:

\be
\begin{array}{lll}
\dot{E}_{1}= A,\\
\dot{E}_{2}=b_j \lambda A,\\
\dot{E}_{3}=c_j \lambda^2 A,
\end{array}
\label{eq:Erate}
\ee

where $A=2 k_{1} \Im[(u^{s_1}_{3} u^{s_2}_{2} u^+_{1})+
(u^{-s_1}_{3} u^{-s_2}_{2} u^-_{1})]$.

As found in \cite{W}, we know that the unstable mode is:

\begin{itemize}
\item  the smallest mode for interactions 1 and 3,
\item  the medium mode for interactions 2 and 4.
\end{itemize}

In order to have a deeper understanding of the energy transfer dynamics,
 we have performed several integrations of eqs.(\ref{eq:triadbis}),
 using the parameters values $\lambda=2$,
 $\alpha=1$, $k_1=2^{-4}$ and different initial conditions.

This analysis, performed on all four models, gives the following
 results:

\begin{itemize}
\item model 1: mode 1 gives energy equally to mode 2 and mode 3,
\item model 2: mode 2 gives more energy to mode 3 and less to mode 1,
\item model 3: mode 1 gives more energy to mode 2 and less to mode 3,
\item model 4: mode 2 gives more energy to mode 1 and less to mode 3.
\end{itemize}

These  energy exchanges are summarized in
 fig.(1).

Behaviors 1 and 4 have already been noticed
by Ditlevsen and Mogensen \cite{DM} for the 3D and 2D GOY
model respectively.

It is also interesting to investigate how these  properties are
 modified when varying the $\alpha$ parameter in the models.
Considering that in eq.(\ref{eq:Erate}) the sum of the
three right-hand sides must
 be zero and
normalizing to one the energy rate on the unstable shell, one can
 evaluate how the
energy sharing between the other shells is affected by changing
 $\alpha$ (see fig.(2)):

 \begin{itemize}

\item In model 1 there is a clear dependence
on $\alpha$. As this parameter
grows up, more and more energy is captured by mode 2.
For $\alpha>1$, the energy gained by mode 2 become greater than the
energy gained by mode 3, leading to a more local energy transfer;

\item  model 3 is
 remarkably independent of $\alpha$, as we shall see in the following.
This fact has very important consequences for the
intermittent dynamics of
the complete shell model;

\item  model 2 has  a
trend analogous to that
of model 1, but with more
 drastic consequences: at $\alpha \sim 1.27$ the mode that receives
 most  of the energy from the unstable mode
 becomes the first  instead of the third-one. This would suggest
 a change in the
direction of the flux: from downward to upward;

\item in  model 4
  the mode that receives  most of the energy remains
 mode 1,  for all values of $\alpha$,
with a consequent reverse energy flux in all cases
 (as it must be, being model 4
 a couple of 2D GOY models).

\end{itemize}

What emerges from this analysis is that
 the behavior of models seems to depend on the choice of the free
parameter $\alpha$, sometimes with strong consequences (as
the direction of the flux in model 2).
The only remarkable exception is the very low dependence of model 3.

Concerning helicity, we can consider the following equations:

\be
\begin{array}{lll}
\dot{H}_{1}(\alpha)= B,\\
\dot{H}_{2}(\alpha)=\eta_2 b_j \lambda^{\alpha+1} B,\\
\dot{H}_{3}(\alpha)=\eta_3 c_j \lambda^{2(\alpha+1)} B,
\end{array}
\label{eq:Hrate}
\ee

where $B=2 k_{1}^{\alpha+1} \Im[(u^{s_1}_{3} u^{s_2}_{2} u^+_{1})-
(u^{-s_1}_{3} u^{-s_2}_{2} u^-_{1})]$ and $\eta_2$
and $\eta_{3}$ depend on the
particular model considered (see table(3)).

By performing the same analysis done for the energy evolution, one can
conclude that helicity is transferred in different ways,
as depicted in
fig.(3).

Being helicity a non-positive defined quantity,
forward (backward) transfer of positive
(negative) helicity is equivalent to backward (forward)
 transfer of negative (positive)
helicity.
In view of this trivial remark,
 arrows in fig.(3) have  only a visual value:
 indicating how helicity (with its own sign) is redistributed
among shells.

Let us notice that models 1 and 3 show a very different
pattern in the helicity exchange among shells.
This can be the explanation of the very different scaling
properties shown by
the two models when varying $\alpha$ (see next section).

For example, we could argue that the dramatic dependence of
the energy exchange on
the $\alpha$ parameter in models 1 and 2, together with the well defined
direction of the helicity transfer, can somehow enhance the role played
by the second invariant with respect to the other two models.

In the next section we will check these
 arguments by performing a numerical study
of eq.(\ref{eq:shells}).

\section{Numerical analysis (model 3).}

In this section we will concentrate on the study of the statistical
properties of model 3 compared with the already known results for the GOY
model (model 1).

Model 3, at difference from model 2, shows for any value of
$\alpha$ a forward energy transfer.

We integrated  eqs.(\ref{eq:shells}) for model 3
 using the standard parameters $\alpha=1$,
 $\lambda=2$, $k_0=2^{-4}$, $f^{\pm}=5(1+i) 10^{-3}$,
 $n_0=1$, $\nu=10^{-7}$ and a total number of shells equal to
$N=22$ and $N=26$.
  In the numerical integration we used
 a fourth-order Runge-Kutta method, with a time
 step varying between $dt=10^{-5}$ (for the simulations with 22 shells)
 and $10^{-6}$ (for the cases with 26 shells).
 Most of the results presented here
 are for the case $N=22$, with a number of iterations of the
order of hundred
 millions, which correspond roughly to several
 thousands of eddy turnover times at the integral scale.
 Stationarity is checked by monitoring the total energy evolution.

  The quantities we have looked at are the structure functions:

\be
S_p(n)=\langle \vert \tilde{u}_n \vert ^p \rangle,
\ee

where $\tilde{u}_n=\sqrt{\vert u^+_n \vert^2 + \vert u^-_n \vert^2}$.

  In fig.(4) we show $\log_2[S_p(n)]$ as functions of
  $\log_2(k_n)$ for
  $p=1,...,8$. There is a well defined inertial range where the structure
  functions follow a power law:

\be
S_p(n) \sim k_n^{-\zeta(p)}.
\ee

  Let us notice that in this model, at variance from the GOY model, there
 are not  the period three oscillations superposed
  on the power-law scaling.
In this case a linear least-square fit allows
  to compute the scaling exponents, $\zeta(p)$,
  with an uncertainty smaller than in the GOY model.

Nevertheless, in order to have a better estimate of the $\zeta(p)$s
 one can study
the moments of a particular
  third-order quantity, the mean energy flux through the $n$th shell:

\begin{eqnarray}
\Pi(n)&=&2 k_n \Im[\langle u^{s_1}_{n+2} u^{s_2}_{n+1} u^+_{n} \rangle +
b_j \langle u^{s_3}_{n+1} u^{+}_{n} u^{s_4}_{n-1} \rangle +
\frac{1}{\lambda} \langle u^{s_1}_{n+1} u^{s_2}_{n} u^+_{n-1} \rangle +
\nonumber\\
      &   &+\langle u^{-s_1}_{n+2} u^{-s_2}_{n+1} u^-_{n} \rangle +
b_j \langle u^{-s_3}_{n+1} u^{-}_{n} u^{-s_4}_{n-1} \rangle +
\frac{1}{\lambda} \langle u^{-s_1}_{n+1} u^{-s_2}_{n} u^-_{n-1} \rangle].
\end{eqnarray}

As pointed out by Pisarenko {\it{et al.}} \cite{PBCFV},
one can write for this quantity
the equivalent of the Kolmogorov's four-fifth law, expressing
the balance between
energy input and energy dissipation in the system.

Considering the energy variation over the first
$n$ shells:
\begin{eqnarray}
\frac{d}{dt}[\sum_{m=1}^n ( \langle \vert u^+_m \vert^2 \rangle +
\langle \vert u^-_m \vert^2 \rangle ]&=&
-2 \nu \sum_{m=1}^n k_m^2 ( \langle \vert u^+_m \vert^2 \rangle +
\langle \vert u^-_m \vert^2 \rangle ) +\nonumber\\
&  &+\Pi(n)+2 \Re[\langle f^+ {u^+_{n_0}}^* \rangle +
\langle f^- {u^-_{n_0}}^* \rangle],
\end{eqnarray}
and assuming a statistical steady state, in the limit of
vanishing viscosity
 we are left
with an inertial range in which $\Pi(n)$ is  constant:
\be
\Pi(n) \sim const.
\ee
In analyzing the scaling properties of all our
results we have always used Extended Self Similarity (ESS). ESS
consists in plotting one structure function versus another ($S_3(n)$,
for example). ESS turned out to improve the precision with which
scaling exponents can be measured in true turbulent flows \cite{BCTBS,BCBC}
and in shell models \cite{BLLP}.

We have applied ESS analysis to two kinds of fit:
\begin{enumerate}
\item $\log_2[S_p(n)]$ vs $\log_2[S_3(n)]$,
\item $\log_2[\Sigma_p(n)]$ vs $\log_2[\Sigma_3(n)]$,
\end{enumerate}
where
\be
\Sigma_p(n)=\langle \vert \Pi(n)/k_n \vert ^{p/3} \rangle.
\ee
In all our simulations we have found the two sets of exponents
coinciding within the
 numerical and statistical errors.

Fig. (5) shows the $\zeta(p)$s
of model 3, compared with those of
the GOY model with the same parameters
 $\alpha=1$ and $\lambda=2$ (corresponding to
the classical choice which conserves the analog of
 the 3D helicity).
The scaling is nearly the same. Indeed,
 we argued in the previous section that both
models have a forward energy flux; what turned out to be different was the
exchange of helicity among shells, together with a different
sensitivity on the
parameter $\alpha$ connected to this second invariant.

What occurs is a strong similarity for $\alpha=1$; nevertheless we expect
a different behavior when this parameter is allowed to vary.

\subsection{The $\alpha$ dependence}

For the $\alpha$ dependence of the  models we have explored two other
different values: $\alpha=0.5$ and $\alpha=2$ (keeping fix $\lambda=2$).

It is well known \cite{BLLP,BKLMW,BK}
that the GOY model shows a strong dependence of its statistical properties
on the $\alpha$ value. For example, if $\alpha < 2/3$ the dynamics is
attracted toward a  fixed point with Kolmogorov scaling, $\zeta(p) = p/3$.
 For $\alpha > 1$ intermittency become more important
than what is usually measured in turbulent flows \cite{BKLMW}.

On the other
hand, the statistical properties of model 3 turn out to be robust under
changes of the $\alpha$ parameter.

In fig.(6) we show the $\zeta_{\alpha}(p)$
exponents for the GOY model and model 3 at $\alpha=0.5,1,2$.
Clearly, there is
an evident dependence of the $\zeta(p)$s on $\alpha$ for the
GOY case while
for the  model 3 the different
exponents coincide within numerical errors.

This behavior is in perfect agreement with the phenomenological
speculations
argued in the previous section from the study of the single-triad system.

The robustness of model 3, with respect to variation
in $\alpha$, gives to this model
 an important
role among the possible shell models of turbulence.

\subsection{The $\lambda$ dependence}

Concerning the dependence on the scale parameter $\lambda$,
 we have performed an exploratory study by
fixing $\alpha=1$ and taking $\lambda=1.5$ and $\lambda=2.5$ in model 3.

For the case
with $\lambda=2.5$ the $\zeta(p)$ exponents are still stuck to the
previous values at
 $\lambda=2$. On the other hand, for the case $\lambda=1.5$ we found
a week discrepancy comparable with the one  found in \cite{BKLMW}
for the GOY model. The issue of what happens in the limit $\lambda
\rightarrow 1$ (the so-called continuum limit)
 is one of the most intriguing problems
that must be analyzed in both cases, GOY model and model 3
(see next section).

In fig.(7) we show three sets of exponents obtained for
 different choices of $\lambda$ for both model 3 and the GOY model.

 The origin and significancy of the weak spreading in the
values of $\zeta(p)$s
 is far from
being understood. By changing $\lambda$, one changes
the ratio between adjacent shells and therefore how viscous and inertial
ranges match together. This non trivial matching maybe
interferes also with  the
determination of the scaling exponents \cite{SKL}.

The $\lambda$ dependence in all this shell models is however
a very important open question due to the obvious interest in having a PDE
describing the continuum limit ($\lambda \rightarrow 1$)
of the energy transfer.
In the following section we derive the equation for the continuum limit
of all four models and we present some proposal
for further investigations.

\section{The continuum limit}

As anticipated in the previous section one of the most interesting and
still unexplored aspect of GOY-like shell models is their
dynamics  in the continuum limit \cite{P}. \\
For continuum limit we intend the limit when the separation between shells
goes to one, i.e. $\lambda \rightarrow 1$:
\be
k_{n+1} = \lambda k_{n} \sim  (1+ \delta) k_{n},
\label{eq:continuo}
\ee
where we have defined  $\lambda = \exp(\delta) \sim
1 + \delta + O(\delta^2)$.
In the limit (\ref{eq:continuo}) we can expand the $u_n$ set as follows:
\be
u(k_{n+m}) = u(k_n) +  \delta\,m\,k_n \partial_k u(k_n) + O(\delta^2).
\label{eq:contshell}
\ee
Taking into account the equivalent  expansions for the $a,b,c$ coefficients
( see table 2) and after  some simple algebra one realizes
that all the three models  (model
1,2 and 3)
lead to the same expression in the continuum, namely:

\begin{eqnarray}
\partial_t u^+(k) &=& i k \left( 4k u^-\partial_k u^+ +
2 k u^+\partial_k u^-
+(2+\alpha) u^+ u^- -\alpha u^- u^- \right)^* + \nonumber \\
 & & -\nu k^2 u^+ +f(k),
\label{eq:cont}
\end{eqnarray}
the corresponding
 equation for the $u^-$s is obtainable from (\ref{eq:cont})
by changing all helicity indeces. \\
This PDE  worths a deeper study for many  reasons.  \\
First, let us notice that the continuum
limit is highly non reversible, i. e.
 trying to come back to a logarithmically-equispaced shell structure
 one does not  recover
the original equations (\ref{eq:shells}). Second, the continuum
model  shows
an unexpected universality: it is the  limit of three models
which have very different behaviors at $\lambda >1$.
 Third, even in the continuum
there are two conserved quantities (in the unforced
and inviscid limit) corresponding to the continuum analogous
of energy and generalized  helicity:
\be
E = \int \frac{dk}{k} (|u^+|^2 + |u^-|^2),\,\,\,
 H_{\alpha} = \int \frac{dk}{k} k^{\alpha}(|u^+|^2 - |u^-|^2),
 \label{eq:enhel}
 \ee
 where the, apparently unusual, $dk/k$ integration step comes
 from the original logarithmically- equispaced shell structure.
 Let us remark
 that the most interesting difference
 between  the continuum expression (\ref{eq:cont})
 and the analog for the old GOY model is that now in (\ref{eq:cont})
  also helicity conservation is well defined. This was not the case
 for the continuum GOY model. The apparent paradox (model 1 is formed
 by two uncorrelated GOY models when  $\lambda > 1$)  is
 easily solved by noticing that in the continuum limit shells collapse
in such a way that the original ordering is destructed. This limiting
procedure introduces a coupling between the two
 sub-models.

This drastic difference with the GOY continuum case suggests
the possibility
that this new set of PDEs has a much richer dynamics than the corresponding
GOY PDEs. In that case, indeed, is quite easy to realize that PDEs are
integrable along the characteristics \cite{L} (at least for the
case of real variables). The solutions have a  burst-like shape with
a Kolmogorov scaling, reaching infinite $k$ at finite time (for zero
viscosity).

Whether eqs. (\ref{eq:cont}) are
more interesting or not is still an open question.

As for the continuum limit of model 4 we need to go the the second order
in the $\delta$ expansion and we obtain:
\begin{eqnarray}
\partial_t u^+(k)& = &i k ( 16 k^2 (\partial_k u^+)^2
 + 2 k^2 u^+ \partial_k^2 u^+
+ 12 (\alpha+2) k u^+ \partial_k u^+ + \nonumber \\
                 &   & +(\alpha+2)^2 (u^+)^2 )^*
-\nu k^2 u^+ +f(k).
\label{eq:cont2d}
\end{eqnarray}
In this case the continuum limit is much more similar to what
should be the continuum limit of a shell model describing 2D turbulence.
The only two conserved quantities are both positive-definite and coincide
with energy and with a generalized enstrophy $\Omega_{\alpha}$. A much more
detailed investigation of both models is  postponed to a forthcoming
study.

\section{Conclusions}

In this paper we have performed a detailed investigation of a
new class of helical-shell models. From the helical-Fourier decomposition
of Navier-Stokes eqs. we have extracted four non-equivalent types
of shell models having two inviscid quadratic invariants similar
to the conserved Navier-Stokes quantities.

Two of these four models coincide
 with the 3D and 2D versions of one of the most interesting
historical shell models:
the GOY model. On the other hand, the other two (model 2 and model 3 in the
text) show different and peculiar properties.
 Most of the numerical results presented in this paper  concern model 3.
 This model revealed
to be much more stable under changes of its free parameters
than  the old GOY model. Why from different (severe)
truncations of  Navier-Stokes eqs.
one ends up with so different dynamical behaviors is certainly
the most stimulating question arising from our study. \\
As for model 2, our preliminary numerical simulations suggest
the presence of a relevant backward energy transfer leading to
possibly strong deviations from the Kolmogorov scaling. A detailed
study of model 2 will be reported later (\cite{BBT}).

 The crucial
role played by  inviscid  invariants
seems to be  confirmed, specially when the phenomenological analysis
suggests the presence of two simultaneous transferred quantities.\\
 Let us notice that usual arguments based on the analysis of
absolute equilibrium behavior \cite{KR,DM} in
order to extrapolate strongly dissipative
effects like intermittency seems to fail  in this class
of shell models. Indeed, starting from the analysis of absolute
equilibrium one should conclude that by changing $\alpha$,
and therefore the
dimension of the second invariant, also the inertial properties in the
dissipative case should change. This is definitely true in the GOY model
but definitely false in model 3.

We have also written down the PDEs describing the semi-universal
(equal for models 1,2 and 3) continuum limit. Investigations
of this PDE are in progress.

Let us conclude with some speculation. Such a rich behavior
of these four models  naturally suggests  that
could be important to ask the same question in the original Navier-Stokes
flow: what are the characteristics of the four sub-models obtained
in the full Navier-Stokes decomposition taking into account
separately the four different helicity classes? Can one
recognize also in the case of  full  3D dynamics sub-models
with important back transfer of energy (like shell model 4
and shell model 2),
or sub-models with different sensibility to the presence of helicity
(like shell model 1 and shell model 3)?

\vspace{1.truecm}

\centerline{ACKNOWLEDGMENTS}
It a great pleasure for us to thank interesting discussion with
D. Lohse, L. Kadanoff,  G. Paladin and  A. Vulpiani. We also warmly thank
Bob  Kerr for having   suggested us to investigate helical-shell models.


\newpage

\centerline{FIGURE CAPTIONS}

\begin{itemize}

\item
Figure (1): Energy exchange in the one-triad system for
 the four models. Dashed (solid) arrows point towards the mode that
 receives less (more) energy.

\item
Figure (2): Variations with $\alpha$ of the energy
rates in the one-triad system for the four models.
For each model are shown the $\dot{E}$ of the two modes
that receive energy from the unstable one, whose energy rate is
always kept equal to 1.\\
a): $\dot{E_2}$ (solid line) and $\dot{E_3}$
(dashed line) vs $\alpha$
for model 1 ($\dot{E}_1=1$);
b): $\dot{E_1}$ (solid line) and $\dot{E_3}$
(dashed line) vs $\alpha$
for model 2 ($\dot{E}_2=1$);
c): $\dot{E_2}$ (solid line) and $\dot{E_3}$
(dashed line) vs $\alpha$
for model 3 ($\dot{E}_1=1$);
d): $\dot{E_1}$ (solid line) and $\dot{E_3}$
(dashed line) vs $\alpha$
for model 4 ($\dot{E}_2=1$).

\item
Figure (3): Helicity exchange in the one-triad system for
 the four models. Dashed (solid) arrows point
 towards the mode that receives less (more) helicity.

\item
Figure (4): The logarithm $\log_2(S_p(n))$
of the structures functions
 of model 3 vs $\log_2(k_n)$. The parameters
values are $\alpha=1$ and $\lambda=2$.

\item
Figure (5): The $\zeta(p)$s of  GOY model
(from \cite{BKLMW}) and model 3. The parameters
values are $\alpha=1$ and $\lambda=2$.
Error bars for data concerning model 3 take into account
both statiscal and power-law fit errors.

\item
Figure (6): The $\zeta(p)$s of GOY model and model 3
for different parameters sets $(\alpha,\lambda)$. $\lambda$ is
always kept equal to 2.
The $\zeta(p)$s ($p=1,..,7$) for the GOY case $(1,2)$
are taken from \cite{BKLMW}.
Notice that for model 3 all data sets collapse on the same curve
for different $\alpha$ values.
Error bars for data concerning model 3 take into account
both statiscal and power-law fit errors.

\item
Figure (7): The $\zeta(p)$s of  GOY model and model 3
for different parameters sets $(\alpha,\lambda)$. $\alpha$ is
always kept equal to 1.
The $\zeta(p)$s ($p=1,..,7$) for the three GOY cases
are taken  from \cite{BKLMW}.
Notice that for one value of $\lambda$ ($\lambda=1.5$) our numerically
evaluated $\zeta(p)$s are slightly different from those found in the
GOY model and in model 3 for other $\lambda$ values.
We are not able to conclude whether this discrepancy is important or not.
Error bars take into account
both statiscal and power-law fit errors.

\end{itemize}


\newpage

\begin{table}
\begin{center}
\begin{tabular}{c|c|c|c|c|c|c|}

 & $s_1$ & $s_2$ & $s_3$ & $s_4$ & $s_5$ & $s_6$\\
\hline
1 & $+$ & $-$ & $-$ & $-$ & $-$ & $+$\\
\hline
2 & $-$ & $-$ & $+$ & $-$ & $+$ & $-$\\
\hline
3 & $-$ & $+$ & $-$ & $+$ & $-$ & $-$\\
\hline
4 & $+$ & $+$ & $+$ & $+$ & $+$ & $+$\\
\hline

\end{tabular}
\caption{Helicity indeces of equations
(\protect\ref{eq:shells}) for the four models.}
\end{center}
\end{table}

\begin{table}
\begin{center}
\begin{tabular}{c|c|c| }

 & $b$ & $c$ \\
\hline
1 & $\frac{\lambda^{-\alpha}-\lambda^\alpha}
{\lambda^{\alpha+1}+\lambda}$ &
$ \frac{-\lambda^{-1}-\lambda^{-\alpha-1}}{\lambda^{\alpha+1}+\lambda} $\\
\hline
2 & $\frac{\lambda^{-\alpha}+\lambda^\alpha}
{-\lambda^{\alpha+1}+\lambda}$ &
$ \frac{-\lambda^{-1}-\lambda^{-\alpha-1}}{-\lambda^{\alpha+1}+\lambda} $\\
\hline
3 & $\frac{-\lambda^{-\alpha}-\lambda^\alpha}
{\lambda^{\alpha+1}+\lambda}$ &
$ \frac{-\lambda^{-1}+\lambda^{-\alpha-1}}{\lambda^{\alpha+1}+\lambda} $\\
\hline
4 & $\frac{\lambda^{-\alpha}-\lambda^\alpha}
{\lambda^{\alpha+1}-\lambda}$ &
$ \frac{+\lambda^{-1}-\lambda^{-\alpha-1}}{\lambda^{\alpha+1}-\lambda} $\\
\hline

\end{tabular}
\caption{Coefficients of equations
(\protect\ref{eq:shells}) for the four models.}
\end{center}
\end{table}

\begin{table}
\begin{center}
\begin{tabular}{c|c|c|}

 & $\eta_2$ & $\eta_3$\\
\hline
1 & $ -1$ & $+1$\\
\hline
2 & $-1$ &$ -1$\\
\hline
3 & $ +1$ & $-1 $\\
\hline
4 & $+1 $& $+1$\\
\hline

\end{tabular}
\caption{Factors in the equations (\protect\ref{eq:Hrate})
for the four models.}
\end{center}
\end{table}
\end{document}